%% file: 0-manuscript.tex
\documentclass[manuscript,screen,authorversion,nonacm]{acmart}

\AtBeginDocument{%
  \providecommand\BibTeX{{%
    \normalfont B\kern-0.5em{\scshape i\kern-0.25em b}\kern-0.8em\TeX}}}

\setcopyright{acmlicensed}
\copyrightyear{2018}
\acmYear{2018}
\acmDOI{XXXXXXX.XXXXXXX}

\acmConference[IDC '24]{Make sure to enter the correct
  conference title from your rights confirmation emai}{June 17--20, 2024}{Delft, Netherlands}

\acmBooktitle{IDC '24: ACM Interaction Design and Children Conference,
 June 17--20, 2024, Delft, Netherlands} 
\acmISBN{978-1-4503-XXXX-X/18/06}

\usepackage{graphicx}

\acmSubmissionID{2559}

\begin{document}

\title[Revitalizing Sex Education for Chinese Children]{Revitalizing Sex Education for Chinese Children: A Formative Study}

\author{Kyrie Zhixuan Zhou}
\thanks{Authors' addresses: Kyrie Zhixuan Zhou, zz78@illinois.edu, University of Illinois at Urbana-Champaign, Champaign, IL, United States; Yilin Zhu, yilinzhu@andrew.cmu.edu, Carnegie Mellon University, Pittsburgh, PA, United States; Jingwen Shan, jshan10@illinois.edu, University of Illinois at Urbana-Champaign, Champaign, IL, United States; Madelyn Rose Sanfilippo, madelyns@illinois.edu, University of Illinois at Urbana-Champaign, Champaign, IL, United States; Hee Rin Lee, heerin@msu.edu, Michigan State University, East Lansing, MI, United States.}
\email{zz78@illinois.edu}
\affiliation{%
  \institution{University of Illinois at Urbana-Champaign}
  \city{Champaign}
  \state{Illinois}
  \country{United States}
}

\author{Yilin Zhu}
\email{yilinzhu@andrew.cmu.edu}
\affiliation{%
  \institution{Carnegie Mellon University}
  \city{Pittsburgh}
  \state{Pennsylvania}
  \country{United States}}

\author{Jingwen Shan}
\email{jshan10@illinois.edu}
\affiliation{%
  \institution{University of Illinois at Urbana-Champaign}
  \city{Champaign}
  \state{Illinois}
  \country{United States}
}

\author{Madelyn Rose Sanfilippo}
\email{madelyns@illinois.edu}
\affiliation{%
  \institution{University of Illinois at Urbana-Champaign}
  \city{Champaign}
  \state{Illinois}
  \country{United States}
}

\author{Hee Rin Lee}
\email{heerin@msu.edu}
\affiliation{%
  \institution{Michigan State University}
  \city{East Lansing}
  \state{Michigan}
  \country{United States}
}

\renewcommand{\shortauthors}{Zhou and Zhu, et al.}

\begin{abstract}
  Sex education helps children obtain knowledge and awareness of sexuality, and protects them against sexually transmitted diseases, pregnancy, and sexual abuse. 
  Sex education is not well taught to children in China -- 
  % leading to a lack of sex-related knowledge and awareness, and an increased rate of pregnancy and sexual abuse. 
  both school-based education and parental communication on this topic are limited. 
  To interrogate the status quo of sex education in China and explore suitable interventions, we conducted a series of formative studies including interviews and social media analysis. 
  Multiple stakeholders such as children, parents, education practitioners, and the general public were engaged for an in-depth understanding of their unique needs regarding teaching and learning sex education. 
  We found that school-based sex education for Chinese children was currently insufficient and restrictive. 
  Involving parents in sex education posed several challenges, such as a lack of sexuality and pedagogy knowledge, and embarrassment in initiating sex education conversations.
  Culture and politics were major hurdles to effective sex education.
  % Games were a favored format to deliver sex education engagingly and acceptably in conservative cultures, yet have not been widely designed and adopted in this context. 
  Based on the findings, we reflect on the complex interactions between culture, politics, education policy, and pedagogy, and discuss situated design of sex education in broader cultural and social contexts. 
\end{abstract}

%% http://dl.acm.org/ccs.cfm.




\maketitle

\input{1-introduction}
\input{2-related}
\input{3-children}
\input{4-method}
\input{5-results}
\input{6-discussion}

\input{7-conclusion}

% \begin{acks}
% \end{acks}

\bibliographystyle{ACM-Reference-Format}
\bibliography{8-references}

\appendix

\input{9-interview}

\end{document}

%% file: 1-introduction.tex
\section{Introduction}

Sex education is important for children in many aspects, for example: (1) obtaining appropriate information about bodies, reproduction, and relationships, (2) establishing healthy relationships, (3) preventing abuses, unintended pregnancies, and sexually transmitted infections (STIs), and (4) enhancing awareness of gender equality \cite{sex:education:1, sex:education:2, sex:education:4, sex:education:6}. 

Sex education for children is currently lacking in China, for several reasons. 
First, like many other Asian cultures \cite{sharma2020vital,joodaki2020ethical,chu2015promoting}, Chinese \textit{culture} is relatively conservative regarding sex, which is an embarrassing, inappropriate, or even taboo topic in families, schools, and society \cite{sex:education:8,qin2023sex,steinhauer2016sex}. 
Second, the \textit{government} puts little effort into promoting people's, especially women's sex well-being \cite{tang2002social}, and even prohibits LGBTQ activism \cite{wang2023fare} -- these are important topics in a comprehensive sex education. 
Third, the Chinese education system puts an extreme emphasis on grades and college admission \cite{novice:teacher}. 
Sex education is not a part of the evaluation, and may thus be deemed ``unnecessary''.

Sex education is not explicitly arranged into the K-12 curriculum in China, but only briefly mentioned across other courses, such as Psychology, Moral Education, Biology, Hygiene, and Physical Education (PE) \cite{situation,rural}. 
As a result: only a small portion of the lesson hours are assigned to sex education; 
the effect of sex education is hardly accountable since it is not evaluated in exams. 
Teachers are not always supported by Education Bureaus and schools, leading to insufficient funding to support their curriculum-developing efforts \cite{rural}. 
The lack of specialized sex education teachers and appropriate teaching materials and resources has resulted in many young people not receiving comprehensive sex education, especially in less developed rural regions \cite{rural}. 
Due to political interference, sex education books discussing LGBTQ topics have been banned \cite{ban}.

Even if parents intend to provide sex education to complement the inadequate sex education provided by schools, they may be too embarrassed to discuss sex-related topics with their children due to their relatively traditional mindset regarding sex and the social stigma around this topic in Chinese culture \cite{sex:education:10}. 
Parents may also encounter difficulty finding sex education materials \cite{sex:education:6}. 
Sex education is particularly lacking for children in rural areas, where people have fewer educational resources \cite{rural}. 

The lack of sex education for children has led to severe consequences. 
Low levels of sexual knowledge lead to risk behaviors in Chinese adolescents \cite{lyu2020sexual}. 
In the years 2014-2016, minors were guilty of 9.46\% of all sexual assault crimes in China \cite{minor}. 
The situation in rural areas is even worse -- according to a research report \cite{lvya}, 17.1\% of the 17,966 surveyed undergrad students had been sexually abused in childhood, and 56.5\% of the victims were female.

In response to the urgent need for sex education for Chinese children in both schools and families, we conducted a series of formative studies to probe the experiences and perceptions of multiple stakeholders including parents, children, education practitioners, and the general public regarding sex education. 
Our goal was to understand the status quo of sex education in China and propose design and policy interventions. 
In particular, we conducted (1) interviews with children and young adults (N=4), parents (N=4), and education practitioners (N=3), and (2) a qualitative content analysis of 333 social media posts from March 2020 to December 2023. Article 1 of the United Nations Convention on the Rights of the Child defines ``children'' as persons up to the age of 18 \cite{detrick1999commentary}. 
In this paper, we use this definition of children and broadly consider sex education in families and K-12 education.

Throughout the study, school-based sex education was found insufficient and restrictive. 
Even if parents deemed it necessary to teach sexual knowledge to their children, they often felt uncomfortable initiating sex education conversations due to cultural taboos, or lacked appropriate knowledge on this topic. 
Our participants expressed expected topics, approaches, and characteristics of sex education. 
For example, games were a preferred way of delivering sex education playfully and casually. 
We conclude by discussing sex education in school, home, and broader cultural, infrastructural, and social settings. 
We contribute by (1) interrogating the status quo of sex education in China, especially challenges and the intertwined relationship between culture, politics, education policy, and sex education, and (2) proposing interventions to revitalize sex education for Chinese children. 

\begin{figure}[htb]
\includegraphics[width=10cm]{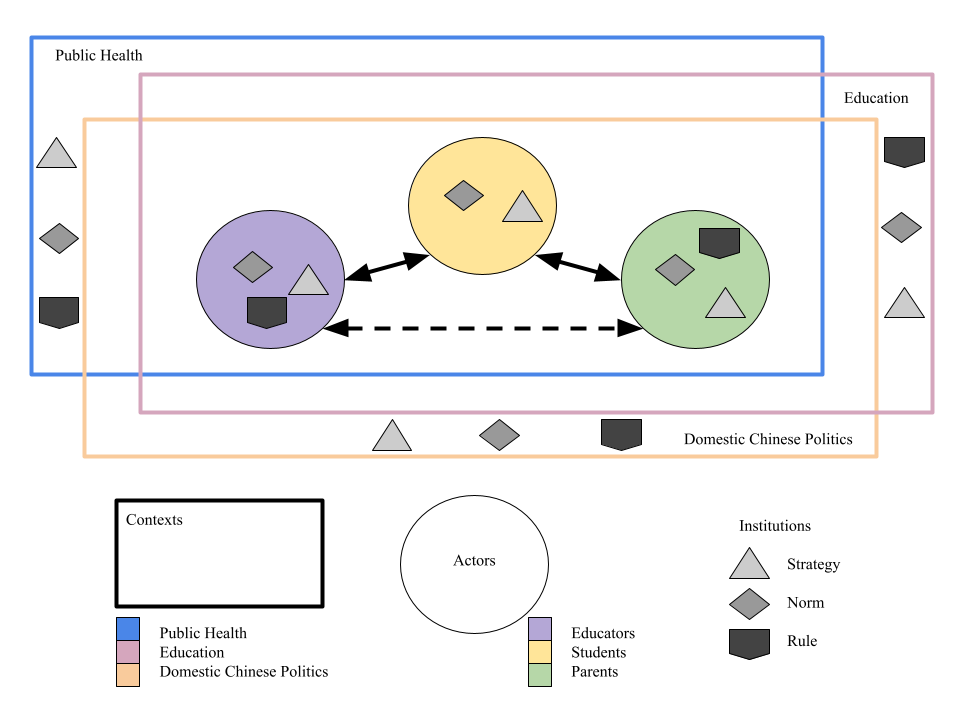}
\caption{Sociotechnical assemblages around Chinese children's sex education.}
\label{fig:F1}
\end{figure}

%% file: 2-related.tex
\section{Related Work}
Sex education spans a wide range of topics, such as the teaching of intimate relationships, human sexual anatomy, sexual reproduction, sexually transmitted infections, sexual activity, sexual orientation, gender identity, abstinence, contraception, and reproductive rights and responsibilities \cite{sex:education:2}. Sex education helps children make informed, positive, and safe choices about relationships, sexual activity, reproductive health, etc. Sex education effectively lowered the risk of adolescent pregnancy, HIV, and sexually transmitted infections \cite{sex:education:2}. Sex education also provides a framework and context for educating students about sexual abuse \cite{sexual:violence}. Below, we discuss sex education in Western contexts and China, respectively, and the role of HCI in sex education.

\subsection{Sex Education in Western Contexts}

Sex education, either school-based or family-based, has been more often researched in Western culture, such as the US \cite{hall2016state} and EU countries \cite{cunha2021integration}. 
The governments have attempted to promote children's rights and health through school-based sex education. 
For example, the Education (No 2) Act 1986 in the UK shifted control to school governors and head teachers and enhanced accountability to parents \cite{des1986education}. 
The public tends to support the teaching of sex education -- in a survey study conducted in 2005-2006 with US adults (N = 1,096), 82\% of the respondents indicated support for sex education programs. 
Sex education has also reached vulnerable populations such as people with disabilities, with additional challenges arising \cite{brown2020design,michielsen2021barriers}. 
Barbareschi and Wu argued that sex education resources were not inclusive to people with disabilities -- communication and consent should be emphasized more since people with disabilities were more vulnerable to sexual exploitation \cite{barbareschi2022accessing}.

Even though school-based sex education is delivered, supported, and even legislated in these countries, researchers have found it insufficient and problematic \cite{cunha2021integration}.  
Kantor and Lindberg argued that the teaching of sex education in the US ``focused almost exclusively on avoiding unintended pregnancy and sexually transmitted diseases, overlooking other critical topics such as the information and skills needed to form healthy relationships and content related to sexual pleasure'', guided by the risk-reduction framework \cite{kantor2020pleasure}.
In the US, abstinence-only, state-level sex education 
policies were found less effective than comprehensive policies, evidenced by a higher rate of sexually active youth and a lower rate of contraception use when youth were sexually active \cite{atkins2021effect}. 
Sex education needs to pay more attention to constructs of heterosexuality and gender norms since teenagers and young adults were found to accept and rationalize controlling and coercive behaviors as ``love'', ``care'', and ``protection'' in heterosexual relationships \cite{abbott2021everyone}. 
LGBTQ young people felt excluded in the sex education classroom in UK schools, which may lead to concrete harm such as poor sexual health literacy, and increased risk of abusive relationships and sexually transmitted infections \cite{epps2023rapid}. 
School-based sex education also faces pressure from parents and politicians who resist the discussion of sex-related topics in the classroom \cite{sex:education:3}.

Family-based sex education is irreplaceable in that parents can establish open communication with children at a young age, and respond to their ideas progressively \cite{sex:education:7}. 
However, family-based sex education also has its limitations. 
For example, mothers tended to be the main educators \cite{sex:education:4}, though fathers' contribution to children's sex education was also deemed important \cite{sex:education:7}. 
Constrained by limited domain knowledge and resources, parents may experience difficulty in providing appropriate sex education -- training for parents and teachers was called for as a solution \cite{kakavoulis2001family}. 
Gender and sib-ship also affect sex education -- boys are less likely to talk to parents than girls, and growing up with a same-gender sibling is associated with lower communication with parents about intimate matters \cite{pasqualini2020parent}. 

\subsection{Sex Education in China}
Sex education in relatively conservative cultures witnesses more challenges such as tensions between sex education and cultural/religious affairs \cite{joodaki2020ethical}.
% (e.g., the necessity of decreasing the age of marriage in Iran)
The topic of sex is taboo in India; sex education was designed by the national government, but was resisted by parents, schools, and states \cite{sharma2020vital}.

Sex education has a long history in China -- Lu Xun (1881-1936), the great Chinese thinker and writer, instructed his students on the subject of sex education back in 1909 \cite{dalin1994development}. In recent years, sex education in China has a similar focus to its counterpart in the US \cite{kantor2020pleasure}, i.e., prevention of pregnancy, unprotected sex, and HIV/AIDS  \cite{aresu2009sex,gao2001aids}. Existing sex education programs have been shown effective \cite{wang2007impact,wang2005potential}. For example,  participation in a sex education program in suburban Shanghai was associated with reduced odds of coerced sex in youth, and increased odds of contraceptive use and condom use \cite{wang2005potential}.

Cultures and politics have a profound impact on the teaching and learning of sex education in China \cite{qin2023sex}. Traditional values and norms in Chinese society such as heterosexual marriages, strict gender roles, childbearing, strong parental roles with little independence of the children \cite{undp2014being}, and Confucian philosophies (e.g., respect, rank/hierarchy, and collectivism) are major barriers to comprehensive sex education \cite{steinhauer2016sex}. Further, the government fails to promote people's sexual well-being or protect vulnerable genders: authorities are known to be indifferent about rape and sexual abuse against women, and women are constructed as legitimate victims of violence \cite{tang2002social}; LGBTQ activism has been strictly prohibited in recent years \cite{wang2023fare}; people's reproduction rights are constantly threatened by China's evolving family planning policies \cite{attane2002china}. These cultural, social, and political factors prevent school- and family-based sex education from being permissive and comprehensive by influencing adults' mindset toward sex, directly banning sex education materials, and so on, which we will elaborate on below.

School-based sex education enhances adolescents’ sexual knowledge and gender awareness and empowers young women \cite{sa2021evidence}. Researchers from Beijing Normal University have devoted great efforts to developing a school-based sex education curriculum and textbooks for teaching primary school kids, training teachers, and involving parents \cite{liu2014school}. However, to date, school-based education is still scarce and non-systematic nationwide, due to several reasons. First, school-based sex education has been constantly impacted by political factors. For example, sex education books used in schools regarding LGBTQ have been banned in recent years due to the government's homophobic sentiments \cite{ban}. 
% There is a relative dearth of research on sex education for sexual minorities (e.g., lesbian, gay, bisexual, transgender, etc.). Understanding the sexual health needs of these groups and their experiences in sex education is critical to improving the effectiveness of inclusive and diverse sex education. 
In response to the censorship of sex education, education materials such as picture books legitimated sex education by decreasing directness in discussions about sex-related knowledge and conforming to heterosexual moralities that pervade mainstream Chinese society \cite{liang2019legitimating}. 
Second, in the examination-oriented education system in China, only examined subjects were emphasized in schools \cite{novice:teacher}, leaving sex education outside the scope of the school curriculum.
% There are hardly dedicated faculty members trained to teach sex education in schools.

Parents in developing countries tend to see sexuality as dangerous, unpleasant, and unsavory and thus are reluctant to talk with their children about it \cite{sex:education:6}. 
The involvement of Chinese parents in the sex education of their children relates to cultural values and family relationship variables \cite{liu2015chinese}. 
According to a large-scale survey in China, only 40\% of parents reported talking to their children about sexuality comfortably \cite{parent}. 
When parents are asked the question ``Where did I come from?'' by their children, they are embarrassed to talk about sex and fertility, but instead, respond with ``You were picked up from the dumpsters'', ``You jumped out from the stone clefts'', ``You used to be a little monkey who lost its tail and became a little child'', etc. \cite{yameng2016come}. 
At the familial and social levels, boys are not considered as important sex education objects as girls \cite{shi2022gender}. 

The current sex education fails to meet the actual learning needs of Chinese children. Limited sexual knowledge contributes to risk behaviors of Chinese adolescents \cite{lyu2020sexual}. Some of them turned to social media, such as WeChat \cite{wang2020social}, as an alternative information source, where inaccurate information exists \cite{social:media}. 

\subsection{Sex Education and HCI}

HCI plays a vital role in promoting sex well-being \cite{sex:education:12,sex:education:13}. 
% For example, online sexual health-related language was analyzed to provide insight into HIV epidemiology \cite{sex:education:12}. ``Protibadi'' was a web and mobile phone-based application designed to report women's stories around sexual harassment in public places \cite{sex:education:13}. 
Efforts have been devoted to developing HCI technologies for sex education. Lin et al. adopted a participatory design approach to help women with minimal sex education backgrounds to better understand menstruation \cite{lin2022investigating}. Games and gamification are commonly adopted to make sex education more engaging \cite{kashibuchi2001educational,kwan2015making,brown2012tackling,gilliam2016lifechanger}. 
Wood et al. developed a game to encourage healthy sexual discussions among adolescents \cite{education:game:3}. 
Arnab et al. developed a digital game to help eliminate coercion and pressure in adolescent relationships \cite{education:game:4}. 
Empirical evidence has shown that gamified sex education was more effective than traditional teaching methods \cite{haruna2021gamifying}. 
Game-based learning with virtual escape rooms motivated and helped students to learn sex education in biology classes \cite{von2022digital}. 
Communal discussions and debriefing are encouraged to maximize the educational effect of serious games in classroom settings \cite{arnab2013development}. 

Culture is an important factor when designing education games for sex education. 
Cao designed sex education games for Chinese immigrants (both children and parents) in Italy, who are influenced by two distinct cultures \cite{cao2017wonderland}. 
Chu et al. believe their sex education game evaluated with Hong Kong teenagers has the potential to be adopted in other regions where discussions on sex are taboo, such as Mainland China, Macau, and Taiwan \cite{chu2015promoting}. 
How children in China utilize interactive systems and games to learn sex education, however, is under-investigated so far.

\subsection{Research Gaps}
Several research gaps exist in the literature on sex education in China. 
First, although games and other HCI techniques have been adopted to teach sex education in Western contexts \cite{kashibuchi2001educational,kwan2015making,brown2012tackling,gilliam2016lifechanger}, if and how Chinese children leverage these novel tools to learn sex education remains unknown.
Second, few studies have explored the experiences and perceptions of multiple stakeholders, including children, parents, and educational practitioners, regarding sex education. 
To have a more situated understanding of the challenges in the teaching of sex education in conservative cultures, particularly China, and inform future education interventions, we present the current formative study to collect diverse data and insights.

%% file: 3-children.tex
\section{Selection and Participation of Children}
Our formative study consisted of two parts: (1) semi-structured interviews with children and young adults (N=4), parents (N=4), and education practitioners (N=3), and (2) a qualitative content analysis of 333 social media posts. We thought it important to collect insights from children to inform more relevant and playful designs of sex education. Thus, children were included in the interviews, being asked about their experiences and perceptions of sex education. We obtained verbal consent from their parents and verbal assent from the children before they entered the interviews.

The authors responsible for collecting and analyzing the user data attended a university in China where no formal IRB processes were available. The remaining authors did not have access to the raw data. Nevertheless, we made all efforts to store the data securely and privately, and reported the results anonymously, to protect the personal information of the children and adult participants.

%% file: 4-method.tex
\section{Methodology}

\subsection{Interviews}

We adopted semi-structured interviews to probe the information and design needs of both adults and children regarding sex education. We specifically asked about how sex education was taught in schools and families, as well as topics in, approaches to, and people's attitudes toward sex education. Example questions included: ``When were you first exposed to sex education? In which way?'' ``When you ask sex-related questions to your parents, what's their reaction?'' The interview questions are included in Section \ref{interview:protocol} in the Appendix.

We recruited participants through a combination of personal contacts and a snowball sampling approach \cite{snowball}. We received few responses, which may be attributed to the discomfort surrounding discussions about sex in Chinese culture. Nevertheless, we strove to recruit 4 parents (P1-P4), 4 children/young adults (C1-C2: children; Y1-Y2: young adults), and 3 teachers (T1-T3). The demographic information of the participants can be seen in Table~\ref{table:1}. All interviewees were from relatively developed regions in Eastern China and had a balanced gender representation. 

The voluntary interviews conducted either in-person or via Zoom lasted no more than 1 hour in early 2020. One interview lasted for only 9 minutes, showing the participant's (P2) lack of knowledge and interest in sex education. We obtained verbal consent from the adult participants. For children, we obtained verbal assent from them and verbal consent from their parents before the interviews began.

We manually transcribed the interviews and adopted a thematic coding approach for analysis \cite{thematic}. 
Two authors independently conducted open inductive coding on the interview transcripts and regularly discussed to reach a consensus. 
XMind, a mind-mapping tool, was used to organize the different levels of themes and quotes into a hierarchical structure. 
Emerging themes included school-based sex education, family-based sex education, and expected features of sex education.  
Below, we use anonymized quotes to illustrate the findings. 

\begin{table}
\caption{Demographic information of the interviewees. P1-P4: parents. C1-C2: children. Y1-Y2: young adults. T1-T3: education practitioners.}
\label{table:1}
\centering
\small
\begin{tabular}{llllll}
\textbf{Alias} & \textbf{Gender} & \textbf{Age} & \textbf{Occupation} & \textbf{Age of Children(s)} & \textbf{Highest Education} \\
P1 & M & 34 & IT Engineer & 6 & Undergraduate \\
P2 & F & 43 & Full-time Housewife & 20 \& 17 & High School\\
P3 & M & 32 & CEO & 2 & Junior High School\\
P4 & F & 30 & University Assistant Professor & 0.5 & PhD \\
C1 & F & 7 & Primary School Student & N/A & Primary School \\
C2 & M & 12 & Primary School Student & N/A & Primary School \\
Y1 & M & 24 & Graduate Student & N/A & Postgraduate \\
Y2 & F & 21 & Undergraduate Student & N/A & Undergraduate\\
T1 & M & 29 & Junior High School Biology Teacher & N/A & Undergraduate\\
T2 & F & 42 & Educational Kid Toy Designer & N/A & Postgraduate \\
T3 & F & 34 & Child Education Specialist & 4 & PhD \\
% \\hline
\end{tabular}

\end{table}

\subsection{Social Media Analysis}
We conducted a social media analysis for two reasons. First, we aimed to generalize our interview results to a larger population given the relatively small sample size and the skewed geographical distribution. Second, we wanted to explore cultural, social, and policy layers, and situate our study within a broader context.

In November 2023, we collected social media posts using the keywords ``children sex education'' on Sina Weibo\footnote{\url{https://weibo.com/login.php}}, a popular Chinese social media platform, using the scraping tool, Bazhuayu\footnote{\url{https://www.bazhuayu.com/}}. A total of 333 posts were collected, ranging from June 2015 to November 2023. On average, the posts were forwarded 45 times, were liked 67 times, and had 16 comments. We applied a similar thematic analysis approach to the posts as in the interview analysis. A wide range of themes emerged, such as the prevalence of sexual harm experienced by children, the importance of sex education, social taboos around discussions on sex, and various approaches to sex education. We elaborate on these aspects with raw quotes from Weibo users.

%% file: 5-results.tex
\section{Findings}

\subsection{Interview Results}
All our participants indicated that sex education hardly existed in China. 
Both school-based and family-based sex education are scarce and have their limitations. 
Y1 sarcastically commented on the insufficient sex education provided by both families and schools, 
\begin{quote}
    \textit{``For many of us, the only way to receive sex education is through porn.''} 
\end{quote}

\subsubsection{School-based Sex Education is Lacking and Restrictive}
Children receive insufficient sex education in schools, as reported by our participants. 
The two primary school students noted that they had no courses dedicated to sex education. 
Parents such as P4 expressed a similar observation:
\begin{quote}
    \textit{``I'm from Shanghai, where people are not conservative. We still don't have much sex education in schools.''}
\end{quote} 

Sex education is not deemed an important topic in K-12 education. Y2 recalled that Physiology classes were arranged in the curriculum when she was in junior high school, but were often taken over by Math or English teachers to give their lectures: 
\begin{quote}
    \textit{``Sometimes the Education Bureau officials go to schools to check if Physiology courses are included in the curriculum as required. At these times, we would pretend to have Physiology classes. After they leave, Physiology classes are taken over by Math or English teachers.''}
\end{quote} 
As we can see, sex education is formal instead of substantial, and is only taught in a way that pretends to comply with policies. This could be attributed to the examination-oriented culture in China, where teachers and parents put an extreme emphasis on students' grades on examined subjects \cite{novice:teacher}.

People in developing countries tend to relegate sexuality to the domain of the dangerous, unpleasant, and unsavory \cite{sex:education:6}. As a result, even if children receive sex education, they are not allowed full access to legitimate sexuality knowledge. Y1 elaborated on the restrictive nature of school-based sex education, where primary school students were only allowed to learn about sexual organs of their own gender in Biology lessons: 
\begin{quote}
    \textit{``I still remember a Biology lesson in 6th grade. We were taught the structures and functions of sexual organs as well as their differences between genders. Boys and girls were taught separately. When the teacher taught about male sexual organs, the girls were asked to stay outside the classroom, and vice versa.''}
\end{quote} 
P3 recalled that when he was in junior high school, the Hygiene course included topics such as personal hygiene and disease prevention, but never touched on sex education. He thought this way of delivering sexuality knowledge was awkward and failed to provide children with a correct and comprehensive understanding of sex. Teachers such as T2 thought the constant avoidance of sex-related topics in children's growth may incur a ``dangerous curiosity'' in them and lead to their misbehavior, 
\begin{quote}
    \textit{``If everyone conceals about sex, it will make children curious and sensitive. Combined with the fact that they don't have correct sex knowledge, they're more likely to do bad things.''}
\end{quote} 

Our participants were overall pessimistic about school-based education. They thought effective school-based sex education was hard to achieve shortly given the heavy emphasis on exams and 
% \textit{Gaokao} 
the National College Entrance Examination \cite{gaokao} in China. T1 explained, 
\begin{quote}
    \textit{``Many parents think Physiology or Hygiene classes are not important at all, and would rather these classes be taken over by Math, Physics or Chemistry teachers.''}
\end{quote} 

\subsubsection{Family-Based Sex Education is Embarrassing and Intimidating.}
Family-based sex education is a natural alternative or complement to school-based sex education. 
However, in a society sanctioning sex conversations, parents tended to feel uncomfortable discussing sex- or body-related topics with their children. 
P4 was a mother who intended to provide sex education to her child but did not know how to start the ``embarrassing'' conversation: 
\begin{quote}
    \textit{``I don't want to neglect sex education, but I just don't know how to start the conversation. I always feel embarrassed when it comes to sex-related topics.''}
\end{quote} 
The parents frequently uttered negative feelings toward sex education such as ``embarrassed'', ``concealed'', ``hesitant'', and ``uncomfortable'', echoing the sexual shame prevalent in Chinese society \cite{sex:education:8}.

Parents' embarrassment in sex education may lead to the teaching of incorrect sexuality knowledge. 
When C2 asked how he was born, his father deliberately skipped the parts about having sex or fertilization in his answer: 
\begin{quote}
    \textit{``I was told by my dad that one needs to get married to have children. Once you get married, you'll automatically have children.''}
\end{quote} 
Such wrong, made-up knowledge has severe consequences -- children will not pay attention to contraception if they do not have legitimate knowledge about pregnancy. Ways to mitigate parents' embarrassment in sex education conversations with their children are, thus, worth exploring. 

Lack of knowledge of sex or pedagogy was another challenge for parents. For instance, P1 acknowledged that he did not know how to teach his son about sex, even if he thought it necessary: 
\begin{quote}
    \textit{``I don't know how to talk to him about it, or where to start. Textbooks are organized in chapters, but sex education is a very broad topic with so many concepts.''}
\end{quote} 
Despite being a university professor with more sexual knowledge compared to other participants, P4 still lacked confidence in delivering the topics to her kid. Mistaken or biased understanding of sexuality was seen among the parents, showing the necessity of co-educating parents and children. For example, P3 wanted their children to have a ``correct'' sex orientation, reflecting the homophobia sentiment among Chinese people \cite{homophobia}. P2 emphasized sex education for girls in particular, echoing the skewed responsibility assignment when it came to sex education and sex crimes \cite{sex:education:11}.

\subsubsection{What Kinds of Sex Education Are Preferred?}
Our participants expressed their opinions regarding what kinds of sex education should be in place for children. 
Gender dynamics was highlighted as an important topic in designing sex education materials and approaches, and the involvement of parents was deemed important in children's exploration of gender dynamics. For example, Y2 wished her mother could join her in fighting for women's sexual liberation: \begin{quote}
    \textit{``I especially hope that my mother can stand on the same front with me, to fight against sexual oppression and to strive for sexual liberation together.''}
\end{quote} 
She further pointed out the unfair fact that girls were taught to protect themselves while boys were hardly taught not to sexually harass others -- she suggested sex education should be provided to all genders. Other important topics to cover in sex education, which were currently missing, included moral discussion about sex, sex desires, safe sex, and sexual orientation, as suggested by the participants.

A gradual, step-by-step approach to sex education is deemed helpful. Y1 noted: 
\begin{quote}
    \textit{``Sex education is intrinsically about education. From the perspective of education, sex education should be staged... We need a complete, staged process of sex education.''}
\end{quote} 
This quote emphasizes the importance of developing educational materials that consider the developmental stages of children, starting with basic sex education and progressing to advanced topics as they grow older. 

None of our participants have heard of or used education techniques, such as games and virtual reality, for teaching and learning sex education. However, games and gamified approaches were repeatedly mentioned as potentially engaging ways to deliver sex education. Further, they served as ideal ice-breakers for parents to initiate the otherwise embarrassing conversation on sex. T3 said, 
\begin{quote}
    \textit{``We must understand what children are thinking about and let them master sex-related knowledge in play''}
\end{quote} 
% Privacy, customization, and staged education were also commonly uttered by the parents as preferred features of sex education.
Features anticipated in such sex education products included being customized for age, privacy, and personalization. 

\subsection{Social Media Analysis Results}
Children are susceptible and vulnerable to sexual harm with limited protection from schools and authorities. Discussions on sex are taboo, effectively making sex education in schools and families restrictive, indirect, and under constant controversy. Although sex education advocates, educators, and initiatives have endeavored to make sex education more acceptable and accessible, parents still feel confused, and sexual crimes/harms persist.

\subsubsection{Sex Education is Important to Protect Children from Sexual Harms.} Sexual harm targeting or experienced by children, such as pregnancy in adolescence, and sexual assault and abuse were widely discussed, often accompanied by the posters' call for more and better sex education to help children protect themselves. 
Child actors were reported to be sexually assaulted by adults, such as celebrity agents, in the workplace. 
School bullying should be escalated if sexual assault is involved, as suggested by many. 
However, in reality, such incidents were often dismissed casually without severe punishments for the perpetrators, their parents, and the schools that did not provide enough education and protection for children. 
% Such opinions were expressed by state media and their editors-in-chief and thus received wide attention and discussion.

Sex education for both children and parents was deemed important to protect children, for several reasons. 
First, children have to suffer from the life-long impact of sexual abuse and other harms experienced in childhood, and sex education can prevent children from these harms to some extent.
A user called for the China Women's Development Foundation to pay more attention to minors' sex education when they heard a 16-year-old girl had an abortion. 
Sex education may simultaneously impact potential perpetrators and potential victims of school bullying that involves sexual assault, 
\begin{quote}
    \textit{``It is really necessary to have a systematic sex education course for children. It not only lets the perpetrators know that they are committing abominable sexual assault but also lets the victims know that they are being sexually assaulted and molested. The victims can promptly report sexual assault to teachers and parents instead of being harmed for a long time.''}
\end{quote}

Second, given the insufficient legal protection for sexual crime victims and light punishment for perpetrators, starting sex education from a young age seems a rare rescue. A supermarket owner was only detained for 9 days for molesting a 14-year-old girl, which was regarded as a ridiculously light punishment. People on social media tended to expect sex education to be delivered to children when they were young, citing the following statement: 
\begin{quote}
    \textit{``You think sex education is too early for your child, but bad people won’t think your child is too young.''}
\end{quote} 
According to them, young children should learn topics such as basic safety knowledge, gender awareness, and bewaring of harm from both strangers and acquaintances. People also cited research that promoted sex education at a young age, arguing that younger children were more curious about sex and accepting of sex education, 
% \begin{quote}
% \textit{``According to research by child psychologists, the golden period for children’s sex education is when they are 3 to 6 years old. At this stage, they begin to be curious and do not have a sense of privacy about their bodies, so it's easier for them to receive sex education comfortably. They are not shy when talking to their parents about sex, and are more receptive to sex education. Parents should not miss this golden period of sex education for their children.''}
% \end{quote}

Third, adults' lack of correct understanding of sexuality hindered children's sex education and even caused extra harm to the victims. One poster said, 
\begin{quote}
    \textit{``Sex education for adults should be promoted as soon as possible. When adults lack a correct understanding of sex, children do not dare speak out about sexual assault and will be harmed again if they speak up.''}
\end{quote} 
Parents’ behaviors, as well as attitudes toward and handling of events related to their children's sexual development, were opportunities for sex education -- parents should educate themselves adequately to set good examples for children, \begin{quote}
    \textit{``Weaning from breast milk, building a sense of security, sleeping in separate rooms from parents, masturbation in childhood, and so on. These are all manifestations of children's sexual development. Parents have an impact on these developmental stages. They should learn sex education knowledge and practice their own educational wisdom. Parents’ attitudes, words, and deeds on sexual issues are the sex education for their children.''}
\end{quote}

\subsubsection{Social Taboo and Failed Sex Education.}
The Chinese government tries to form an ``asexual reproduction'' culture in society, according to many -- it simultaneously encourages an increase in fertility among women \cite{liu2023fertility} and sanctions sex-related discussions. Such prohibitive governmental attitudes toward sex impacted multiple facets of society. For instance, in the entertainment industry, sex-related words are increasingly being censored, \begin{quote}
    \textit{``In movies and TV shows, `sanitary napkin' is `makeup remover cotton' and `menstruation' is `adolescence'.''}
\end{quote} 
As seen, even words related to women's bodies were censored, demonstrating the heightened level of censorship of sex-related discussion. The social taboos around sex have hindered sex education greatly. One post attributed the failure and difficulty of children's sex education in China to the ``sex-phobic ideology of the entire era and society.'' 

When it comes to \textit{family-based sex education}, parents tend to worry that children would have sexual intercourse at a younger age if they ``know too much about sex'', although one user refuted this argument with research that provided opposite conclusions. Parents also found difficulty when explaining to children how they were born -- many told their children that they were ``picked up from the trash can,'' echoing prior research \cite{yameng2016come}. One person talked about how her parents prohibited her from sex education at all stages of her growth, 
\begin{quote}
    \textit{``When I was in kindergarten, I read a picture book for sex education. There was a scene where a man sexually assaulted a girl. I didn't know what it was and asked my mom. She told me to not read it since I was too young... When I was in junior high school, I saw condoms and asked my parents what they were. They were like, `You will know when you grow up.' ''}
\end{quote} 
The situation was even worse for parents with lower education levels and from less developed areas -- they avoided talking about sex with their children to a larger extent -- leading to a higher rate of sexual assault against children in rural areas, according to sex education advocates on Weibo. Only a few parents shared how they explained sexual organs, pregnancy, and other sex education concepts to their children.

Conservative parents also actively and successfully interfered with \textit{school-based sex education}. Some parents united to report sex education textbooks to the government, which in turn banned the books. Sex education advocates criticized these parents for depriving their children of the opportunities to receive sex education and protect themselves against sexual harm, 
\begin{quote}
    \textit{``The sex education textbooks compiled by Beijing Normal University were really rigorous and professional. After they were used in primary schools in Shanghai, they were reported by the parents and banned. I was heartbroken at that time. These parents should be nailed to the pillar of shame. Do you know how many left-behind children\footnote{Left-behind children in China are children who remain in rural regions of the country while their parents leave to work in urban areas.} rely on sex education textbooks to survive? Families and the government cannot protect them. They can only rely on the books to protect them.''} 
\end{quote}

The failure of sex education has severe consequences. One poster related the high rate of unintended pregnancy in adolescence to the lack of family- and school-based sex education.
% echoed this view, \textit{``Sex has always been a closely guarded topic in China, and children’s sex education has therefore been intentionally ignored for a long time.''} 
Sex education advocates argued that instead of holding prejudice against sex education, parents should overcome their shame and face it calmly; and that comprehensive sex education was hardly accomplishable unless children were treated as independent and respected individuals, instead of ``innocent babies for a lifetime.'' An insightful post wrote,
\begin{quote}
    \textit{``Sexual knowledge is not shameful; sexual ignorance is shameful.''}
\end{quote}

\subsubsection{Approaches to Sex Education.} A variety of approaches to sex education were discussed and promoted. Parents, educators, and advocates recommended educational materials such as documentaries, cartoons, videos, and books/book lists regarding sex education to parents -- sex crimes and inadequate sex education were cited in the recommendations to draw parents' attention. 
A mother shared how her daughter learned about menstruation through educational videos that vividly explained the concept.
A sex education book emphasizing the empowerment of children and discussing a wide range of ``rare topics'' (e.g., LGBTQ, pornography, sex, and pregnancy) was favored by many parents for its readability -- it discussed concrete cases in the form of Q\&A. 
% The parents acknowledged that in the short video era, their attention spans were greatly shortened, making them unable to read tedious books. 

Sex education advocates actively devoted themselves to spreading scientific knowledge (e.g., by delivering online courses), compiling textbooks, developing websites and apps, and calling for sex education. A professor was widely praised by people for her textbook compiling efforts, although the sex education textbooks she developed have been constantly resisted by conservative parents. 
A group of sex education practitioners and researchers have come up with a guide for sex education based on years of research and education practices. 
% The Guide included eight core concepts: interpersonal relationships; values, rights, culture, and media and sex; social gender; violence and safety; health and well-being skills; human body and development; sex and sexuality; and sexual and reproductive health. The Guide provided customized materials for different age groups, with different learning objectives and levels of complexity.
% A sex education teacher taught their students to embrace their gender identities instead of being confined by their biological sex, \textit{``Boys can be masculine. Boys can also be vulnerable. Girls can become wives and mothers, but they can also become whatever they want to be.''} 

Several initiatives worked to provide sex education, curate reading lists, and hold sex education exhibitions for children and parents. Notably, some initiatives specifically provided sex education to children in rural and mountainous areas, or children with disabilities, who had access to less educational resources. For example, the Guarding Flower Buds Project aimed to provide sex education to left-behind girls in rural areas.
In one post, they announced the books that were selected as sex education materials and donated to primary school students, 
% \textit{``[Announcement of Book Selection Results] Recently, we invited children's sex education experts and practitioners, active teachers in rural areas, and book donor representatives to online meetings, discussing the books included in the gift packs. Finally, based on professional suggestions and voting results from all parties, one book was selected for girls in grades 1 to 3 and one book for girls in grades 4 to 6.''}
According to them, the selected books met the actual needs of rural primary schools and were scientific, comprehensive, nice-looking, and understandable.

A sex education website utilized gamification to make sex education more acceptable. However, gamification approaches and educational games were only occasionally seen in the discussion.

%% file: 6-discussion.tex
\section{Discussion}
Children face a wide range of issues regarding sexuality and gender identity, such as sexting among teenagers \cite{sex:education:15} and gender bias \cite{sex:education:16}. 
% Sexting and nude selfies are important topics in high school classrooms as adolescents, especially girls, are at risk of privacy violations 
% Gender bias is everywhere, from fairy tales \cite{sex:education:16}, which are children's regular reading materials, to chatbots \cite{sex:education:17}. 
Through our study with diverse stakeholders, including parents, children, education practitioners, and the general public, we revealed the insufficient and restrictive sex education for Chinese children, echoing prior literature \cite{situation, rural}. 
Based on evidence from our formative studies, we delineate how sex education should be taught and how to unite interventions from multiple stakeholders at various levels to approach this complex issue. 

\subsection{How Should Sex Education be Taught?}

\subsubsection{Starting Sex Education at A Young Age.} 
Sanderjin van der Doef, a Dutch psychologist, said, ``As soon as children have questions, they have the interest, and then they have the right to get a correct answer'' \cite{young:age}. Parents on social media want children to be taught sex education early since children start to face risks of sexual harm at a young age. Substantial evidence has supported sex education beginning in primary school \cite{sex:education:5}. According to Erikson's Stages of Psychosocial Development \cite{orenstein2022eriksons}, early in children's development, they explore things that interest them in the world around them, including their own bodies -- they treat sexual knowledge like any other scientific knowledge such as astronomy and geography without feeling shame. We argue that sex education should be delivered at a younger age in families, especially when school education on sex is largely lacking in China \cite{rural, situation}. 

\subsubsection{Educational Informatics Practices}
Our results highlight the need for pedagogical games and technologically mediated sex education, within the practice of educational informatics and in the context of K-12 education \cite{srivastava2012educational}. Not only do these approaches address the emotional and normative concerns (e.g., taboo and embarrassment around sex) identified in our results, but they also align with broader informatics lessons about the impact of cultural identity and values upon engagement with information and technology ~\cite{Fichman_Sanfilippo_2013}.

Previous research has validated the role of games in providing engaging sex education for children \cite{kashibuchi2001educational,kwan2015making,brown2012tackling,gilliam2016lifechanger}. Sex education games can potentially help mitigate parent embarrassment and engage children in sex education in conservative cultures, as suggested by our interviewees. Educational games are also suitable for educating parents who acknowledged shortened attention spans in the ``short video era'' and indicated difficulty in reading books. However, the use of sex education games was absent in our interviewees and sporadic in social media discussions -- only a few website/app creators promoted their sex education products. Future research should design and evaluate games to teach sex education in conservative cultures. The games may want to adopt the desirable features expressed by our participants, e.g., covering currently missing topics like diverse gender identities, considering children's cognitive development, involving parents, and treating children of all genders as important sex education objects \cite{gender:difference}.

Other advances in HCI technology may also contribute to sex education, such as personalized learning experiences based on students' needs and interests \cite{pratama2023revolutionizing}. Intelligent tutoring could help students set up learning goals, solve problems, provide advice, and improve students' learning outcomes in sex education \cite{lin2023artificial}, which is particularly helpful when dedicated sex education teachers are currently lacking. 
% Through the establishment of a safe and private sharing space, under the premise of respecting students' privacy, it realizes the interaction and exchange of views, learning resources, exchange of ideas on the diversity of sex, and improvement of the learning effect. 
% The sharing and archiving of sex education learning resources through cloud computing technology could encourage resource management and collaborative work to improve learning efficiency.

\subsection{Sex Education in School Settings}
An overhaul is needed to improve school-based sex education in China. Since the education system puts an extreme emphasis on exams \cite{novice:teacher}, there is hardly a dedicated course for sex education in K-12 education. Instead, sex education spreads across other courses such as PE and Biology, and is only occasionally mentioned. 
Teachers of Chinese, Math, and English, the subjects in the National College Entrance Examination, tend to occupy the ``unimportant courses'' such as Hygiene. 
While most schools and teachers choose to deliver examination-oriented education instead of quality-oriented education given the pressure from society, education bureaus, and parents \cite{novice:teacher}, we see a need to emphasize the importance of sex education to teachers. 
The government is encouraged to reconsider its education policy and enforce sex education in schools given the prevalence of sexual crimes targeting and committed by adolescents \cite{minor}. The Education (No 2) Act 1986 in the UK \cite{harris1994education} is a valuable reference point.

\subsection{Sex Education in Home Settings}
Parents are key to their children's sex education \cite{sex:education:7}, especially when the overhaul of school-based education is not likely to take place shortly in China. However, including parents in sex education can be challenging. Sex education is an umbrella of many sexuality-related topics \cite{sex:education:2}, thus educators are required to have a relatively broad knowledge of sexuality to be able to deliver sex education to children. 
Through the literature review (e.g., \cite{sex:education:6}) and our interviews with parents, we found they often lacked basic knowledge of sexuality and access to quality information. 
To extend the ineffective school-based sex education \cite{sex:education:5, sex:education:1, sex:education:9} into the family realm and the younger phase of childhood, co-education of children and parents is necessary through educational games, programs, etc. 
For example, an information page may be included in sex education apps/games to provide parents with a wide range of information sources. 
Future research could also consider studying social media channels for parents to exchange information and knowledge regarding sex education.  

\subsection{Sex Education in Broader Cultural, Infrastructural, and Social Settings}

% \subsubsection{Providing Comprehensive Sex Education in Conservative Cultures.} 

Prior analysis of social factors on contextual norms and practices in China, grounded in Hofstede’s model of cultural dimensions, emphasize face-saving practices and non-confrontational information/communication patterns due to power distance, motivation toward achievement, and uncertainty avoidance~\cite{Pareek_Kumar_2020}. Our results echo these insights and point toward the relative lack of progress, despite evidence supporting the need for sex education and students' demand.

The social stigma around sex topics rooted in Chinese culture \cite{sex:education:8}, combined with political interference, has led to restrictive and ineffective sex education -- children were not allowed access to legitimate sex knowledge in both families and schools, and were poorly prepared to protect themselves from sexual harm. No existing practices guarantee effective teaching of sex education in the Chinese context. Designing culturally situated sex education remains an important and timely research direction \cite{eglash2006culturally}.

\subsubsection{A Whole Picture: Conflicting Strategies and Norms}
Our formative studies highlight existing tensions between pedagogical strategies, evidence-supported norms regarding children’s well-being, conservative social norms regarding sex, and Chinese educational policy. While interactions around sex education in China are limited, they are supported by a variety of information and communication channels, including technological platforms, curriculum, traditional information resources, and games. The current status of sex education in China and barriers to change are a product of: political interference, as a strategy; entrenched social norms discouraging discussion within the sociopolitical context of China; a lack of public health laws requiring sex education; parental inaction due to strategies grounded in indifference and discomfort, as well as norms to defer to educators; and educators who are left to resolve conflicts among these strategies, norms, and rules. This complex reality reflects principles of social informatics regarding sociotechnical assemblages and the interaction effects and dynamics associated with multiple interactions and institutions in networked contexts, as depicted in Figure~\ref{fig:F1}. Future research is encouraged to further interrogate these interactions.

%% file: 7-conclusion.tex
\section{Conclusion}
In this paper, we present a series of formative studies to unearth how sex education is currently delivered to children in China and elicit information and design needs from multiple stakeholders. Sex education is found limited and obscure in both schools and families, arising from the complex interactions between culture, politics, education policy, and pedagogy. 
We discuss sex education in school, home, and broader cultural and social settings, and propose possible interventions. 
% A closer look reveals challenges perceived by adults in initiating sex education conversations with their children, such as embarrassment arising from the social stigma around sex and a lack of comprehensive sexuality knowledge. 
% Based on the findings, we reflect on delivering sex education in conservative cultures and think of ways to deliver more comprehensive sex education for children. 
With this study, we hope to inspire more research in designing and evaluating socially acceptable sex education in conservative cultures. 

%% file: 9-interview.tex
\section{Interview Protocol}
\label{interview:protocol}

\begin{enumerate}
    \item How do you feel about sex education? 
    \item How do you define sex education? 
    \item Have you received sex education from schools? When were you first exposed to sex education? In which way?
    \item Have your parents given you sex education? In which way? Who initiated the conversation, you or your parents? When you ask sex-related questions, what's their reaction?
    \item In which way do you expect schools and parents to convey sex education? What topics do you think should be covered in sex education?
    \item (For parents/teachers) How do you teach your children/students sex education?
    \item Have you used educational games/tools/materials about sex education? What features do you expect in them?
\end{enumerate}